%% file: motag6.tex
\newif\ifTR
\TRtrue

\ifTR
\documentclass[onecolumn,11pt]{IEEEtran}

\else
\documentclass{IEEEtran}
\fi

\usepackage{epsfig,endnotes,url}
\usepackage{comment}
\usepackage{graphicx,subcaption}
\usepackage{amsfonts,amsmath}

\newcommand{\squishlist}{
   \begin{list}{$\bullet$}
    { \setlength{\itemsep}{0pt}      \setlength{\parsep}{0pt}
      \setlength{\topsep}{3pt}       \setlength{\partopsep}{0pt}
      \setlength{\listparindent}{-2pt}
      \setlength{\itemindent}{-5pt}
      \setlength{\leftmargin}{1em} \setlength{\labelwidth}{0em}
      \setlength{\labelsep}{0.5em} } }
\newcommand{\squishend}{
		\end{list}  }

\newcounter{qcounter}
\newcommand{\squishlistnum}{
   \begin{list}{\arabic{qcounter}}{\usecounter{qcounter}}
    { \setlength{\itemsep}{0pt}      \setlength{\parsep}{0pt}
      \setlength{\topsep}{3pt}       \setlength{\partopsep}{0pt}
      \setlength{\listparindent}{-2pt}
      \setlength{\itemindent}{-5pt}
      \setlength{\leftmargin}{1em} \setlength{\labelwidth}{0em}
      \setlength{\labelsep}{0.5em} } }
\newcommand{\squishendnum}{
		\end{list}  }

\input{define}

\newcommand{\qed}{~\\ \raisebox{.65ex}{\fbox{\rule{0mm}{0mm}}} ~\\}
\date{}

\begin{document}


\title{\Large \bf Moving-target 
 Defense against Botnet Reconnaissance\\
and an Adversarial Coupon-Collection Model\thanks{Our
research is supported by Defense Advanced Research Projects Agency (DARPA) Extreme DDoS Defense (XD3) contract no. HR0011-16-C-0055, Amazon AWS and Google GCE credit gifts, and a Cisco Systems URP gift. Opinions, findings, conclusions, and recommendations expressed in this material are those of the authors and do not necessarily reflect the views of DARPA, Amazon, Google, or Cisco.
We thank Jon Thompson of Akamai for a very informative discussion.
}}

\author{
{\rm Daniel Flec$\mbox{k}^1$, George Kesidi$\mbox{s}^2$, 
 Takis Konstantopoulo$\mbox{s}^3$,  
Neda Nasirian$\mbox{i}^2$,} \\
{\rm Yuquan Sha$\mbox{n}^2$,  and Angelos Stavro$\mbox{u}^1$}\\
~\\
\begin{tabular}{ccc}
1. CS Dept, GMU & 2. School of EECS, PSU
 & 3. Dept. of Math, Univ. Liverpool\\
\{dfleck,astavrou\}@gmu.edu & 
\{gik2,nun129,yxs182\}@psu.edu  
& takiskonst@gmail.com
\end{tabular}
}

\maketitle

\begin{abstract}
We consider a cloud based multiserver system consisting of a set of replica
application servers behind a set of proxy (indirection) servers
which interact directly with clients over the Internet.
We study a proactive moving-target defense to thwart a DDoS attacker's 
reconnaissance phase and consequently reduce the attack's impact. 
The defense is effectively a 
moving-target (motag) technique in which the proxies dynamically change. 
The system is evaluated using an AWS prototype of HTTP redirection 
and by numerical evaluations of an ``adversarial"
coupon-collector mathematical model, the latter
allowing larger-scale extrapolations.
\end{abstract}


\section{Introduction}
Two very significant, high-volume botnet based DDoS attacks were
witnessed in Fall 2016 \cite{DDoS-Dyn-102116,Krebs16}. The Dyn attack was launched by IoT-device bots (compromised using factory default credentials) against Oracle DNS service. Since 2016, other significant attacks have involved, \eg
the Mirai, Hajime  and BrickerBot botnets, which largely consist of
IoT devices (compromised typically via known exploits)\footnote{Not all
major DDoS attacks involve IoT based bots, \eg the recent brief
but intense amplification attack targeting 
GitHub used vulnerable Memcached servers
\cite{GitHub-DDoS-18}.}.
Indeed, many (not just legacy) IoT devices cannot be secured.
Thus, infrastructure based defenses need to be mobilized against   
such DDoS attacks.  Such defenses can be situated at the attacker-side
network edge (\eg egress filtering, intrusion detection),
within the network (\eg Akamai Prolexic), or on the premises of
the targeted victim (enterprise based defenses, \eg
reactive dispersive autoscaling,
challenge-response, and anomaly detection).  
In this paper, we consider the last context.

{\bf Threat model considered herein:}
Consider a botnet with roughly two
types of bots: weak (\eg IoT based) and powerful.
Both types can effectively flood 
{\em e.g.} with session requests given the IP 
address (or IP and port number) of the target(s).
Powerful bots can cope with server redirection, perform DNS lookups, and
even mount mock sessions.
Much more numerous weaker (IoT-based) bots  cannot set up mock sessions and may not be able to cope with server redirection.
A few more powerful bots may be used perform reconnaissance and then target the rest.

As the hardware computing platform of an IoT device is typically very limited,
certain (\eg HTTP, RTSP) protocol features are removed if they are
not required for the IoT device's nominal use (this also
potentially reduces vulnerabilities - the attack surface - of the
IoT device). 
A key assumption in the following is that the majority of
the bots required for the flooding attack to be effective 
cannot cope redirection. Clearly, future attacks wherein
the malware itself provides redirection  functionality are possible.

\begin{figure}
\begin{center}
\ifTR
\includegraphics[width=0.75\textwidth]{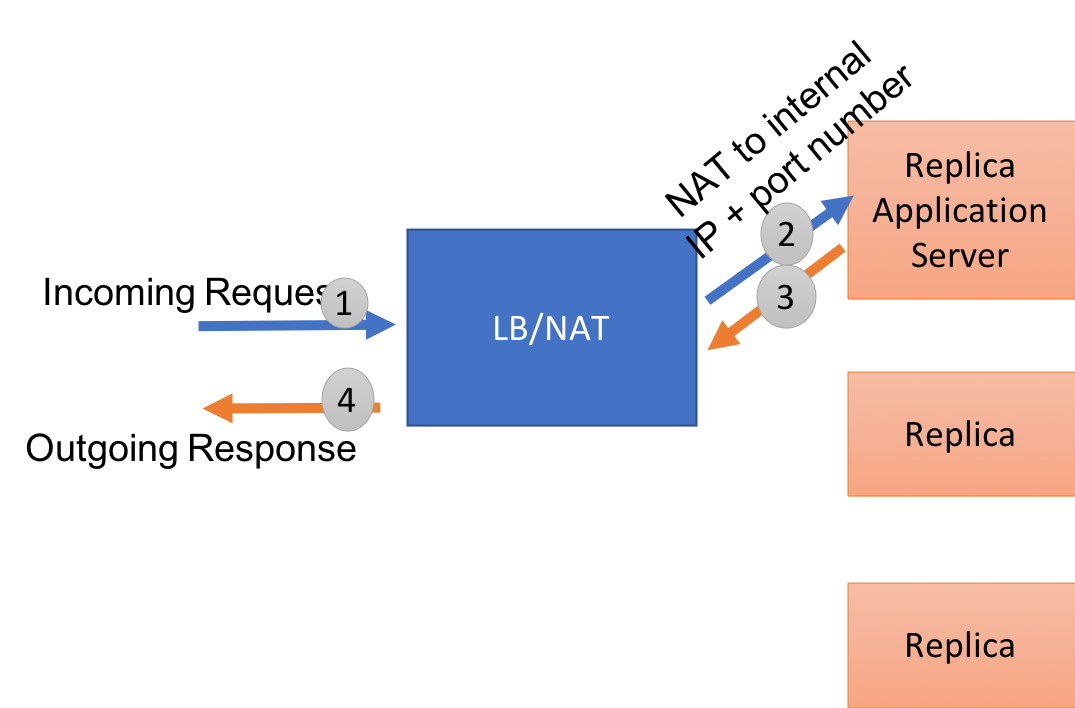}
\else
\includegraphics[width=\columnwidth]{./figure/tenant-baseline.png}
\fi
\caption{Baseline tenant with LB/NAT and no proxies.}\label{fig:tenant-baseline}
\end{center}
\end{figure}

{\bf Attack target:}
First, consider a tenant with a number of replica application servers (replicas).
Its baseline reaction to DDoS attacks or flash crowds is to autoscale the replicas (automatically increase their number).
The server/client loads could be balanced by round-robin DNS or via a Load Balancer (LB).
The LB could 
act as an (inline) Network Address Translator (NAT) or
redirect sessions to the replicas.
A tenant could have multiple LBs via round-robin DNS.
Also, a tenant could have a layer of indirection servers (proxies) between the LB(s) and replicas, and, again, the LB(s) could either NAT,
see Figure \ref{fig:tenant-baseline},
or redirect sessions to the proxies.

\begin{figure}
\begin{center}
\ifTR
\includegraphics[width=0.75\textwidth]{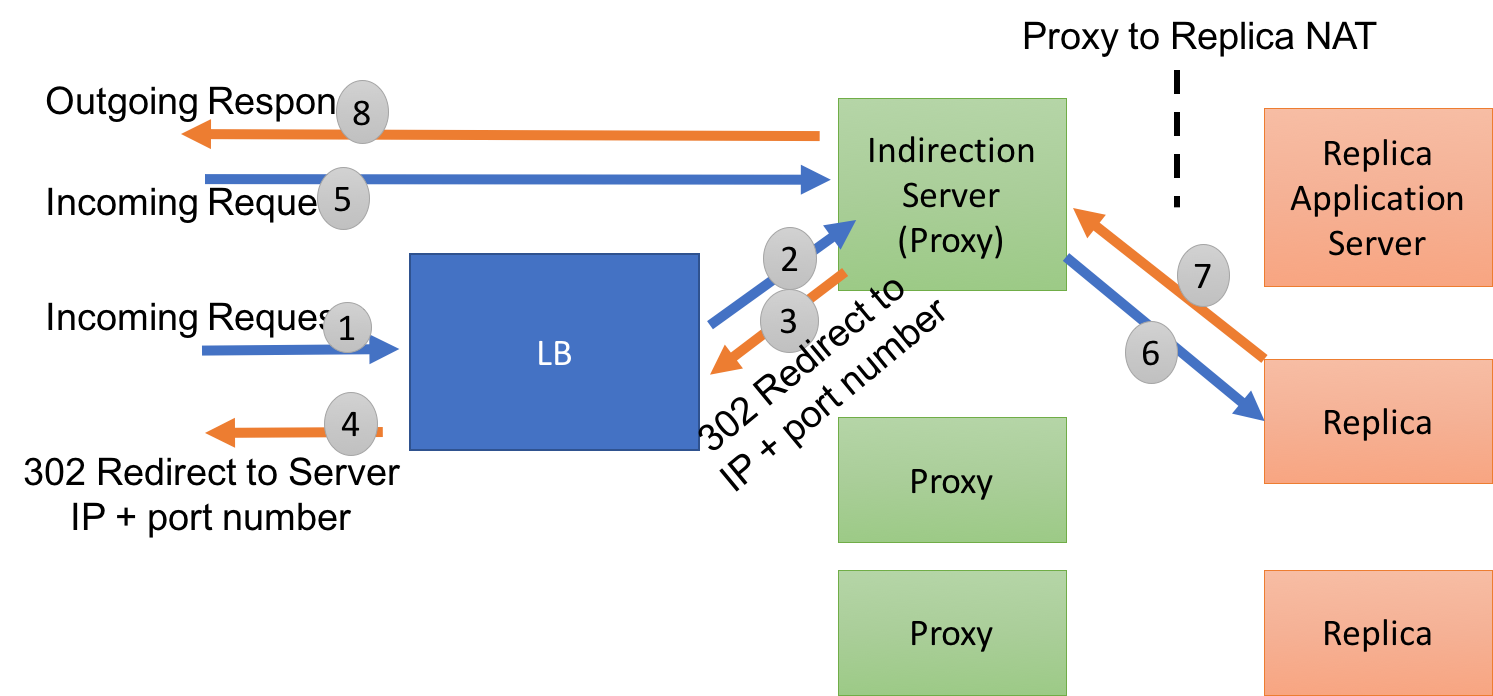}
\else
\includegraphics[width=\columnwidth]{./figure/tenant.png}
\fi
\caption{LB-to-proxy redirection.}\label{fig:tenant}
\end{center}
\end{figure}

We herein consider a tenant that services a plurality of clients with a plurality of replicas.
DNS resolves clients' session requests to the tenant's LB.
The LB redirects session request to a proxy, \ie the LB is not a NAT.
Clients contact their proxies directly and do not know the address of their replica server.
See Figure \ref{fig:tenant}.
The tenant will have a coordination server that
will, in particular, control the DNS records of LBs and proxies,
and the NAT from proxies to replicas.
Load-balanced content distribution  networks (CDNs)
or session establishment 
systems are examples adaptable to this architecture.

Each replica can be reached from plural proxies (by proxy-to-replica NAT).
Though a  volumetric DDoS attack would typically overwhelm 
the proxies before the LB, \ie the LB would not be the bottleneck,
the proxies may shield the replicas\footnote{But
the proxies themselves do not protect against 
``algorithmic" (non-volumetric,  application layer)
DDoS attacks such as Slowloris \cite{Slowloris} and 
BlackNurse \cite{BlackNurse2}. The defense against such
attacks is not considered herein.}.

{\bf LB-to-proxy redirection:}
The proposed defense requires LB-to-proxy session redirection.
LB-to-proxy redirection by proxy domain name requires that clients obtain proxy (IP, port number) by DNS (note that DNS cannot be used to implement NAT).
Otherwise, if redirection is by proxy (IP, port number), then nominal clients need to obtain the proxy's domain name by reverse DNS for HTTPS 
(web servers associate SSL certificates to domain names).
There is
increased attacker workfactor  if redirection is by 
proxy domain name because a bot needs to query DNS 
 before flooding the proxy;
note that flooding bots don't care about SSL certificates.
If the LB redirects to proxies by {\em both} proxy (IP, port number) 
and domain name, then 
additional DNS querying by the client may not be required for 
redirection\footnote{Once such a redirection message is received, the client
would first update its local DNS cache accordingly.}, 
and proxy DNS records would
only need to be stored locally by coordination server and LB;  
but this would be a  change to the HTTP redirection mechanism.
Unlike previous work that has suggested changes to SSL/TLS,
\eg  changes to TLS to support middleboxes \cite{Steenkiste15},
these proposed changes are the minimum required to streamline hot
server migration.

{\bf This paper is organized as follows.}
In Section \ref{sec:defense-overview},
we motivate and describe the proactive moving-target defense.
In Section \ref{sec:cc}, we briefly describe an ``adversarial"
coupon-collection model for botnet reconnaissance of proxies
under moving-target (motag) defense.
In Section \ref{sec:emul}, we describe our AWS emulations
and give example results. These results are compared against 
an adversarial coupon-collection model allowing extrapolation to 
larger-scale scenarios.
After a brief summary in 
in Section \ref{sec:concl} and acknowledgements,
the paper concludes with an Appendix on adversarial
coupon collection models.

\section{Proactive Moving-Target (motag)
Defense against Reconnaissance by Changing 
LBs and Proxies}\label{sec:defense-overview}

\begin{figure}
\begin{center}
\ifTR
\includegraphics[width=0.75\textwidth]{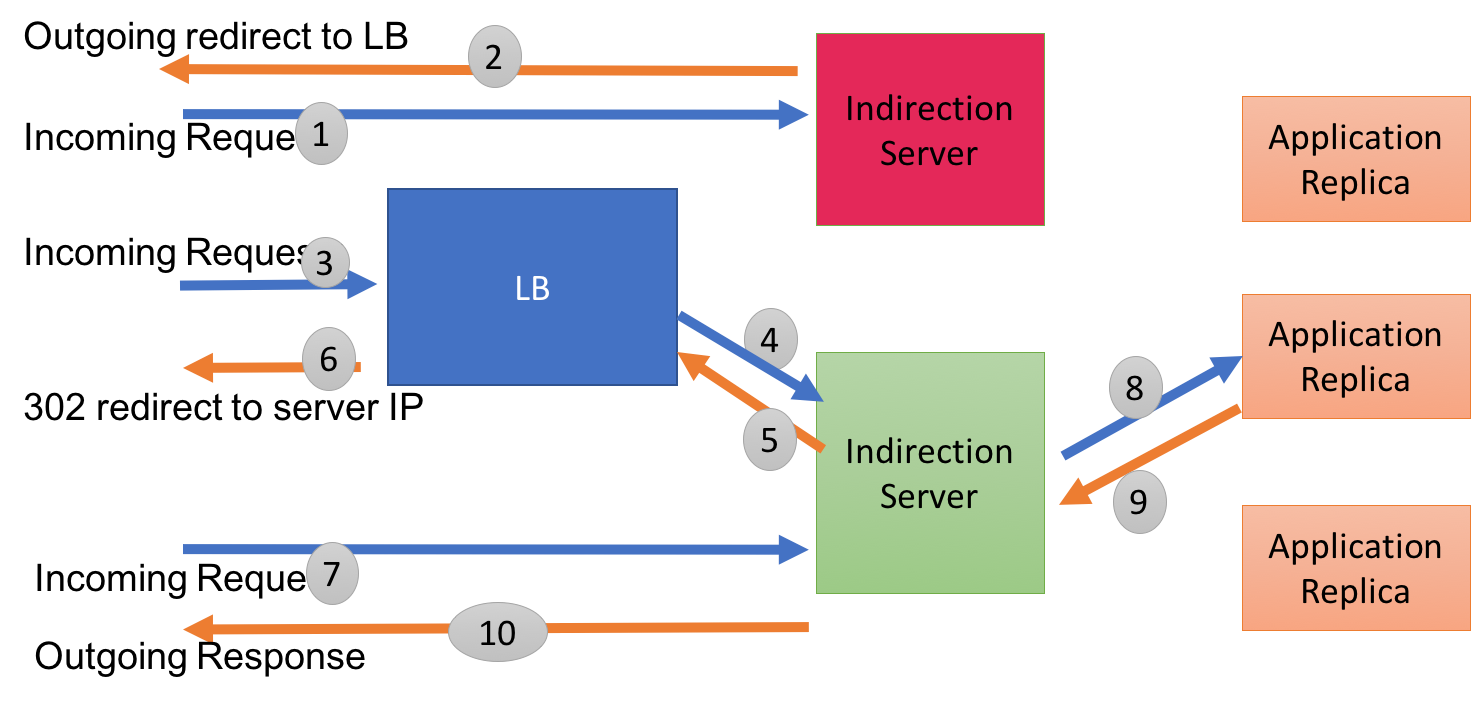}
\else
\includegraphics[width=\columnwidth]{./figure/motag-p2p.png}
\fi
\caption{Proxy-to-proxy redirection.}\label{fig:motag-p2p}
\end{center}
\end{figure}

Under our proposed motag defense,
the LB may change its IP periodically via DNS (if plural LBs under round-robin DNS, then the set of LB IPs periodically change).
Proxies age-out and are replaced by new  ones with different IPs.
Proxy Virtual Machines (VMs) of the same tenant need to be on different physical servers, i.e., different proxies have different IPs  (different VMs on a physical server are discriminated by port numbers).
So, a flood targeting tenant X's proxy may have collateral damage on other 
tenants' VMs sharing the same physical server, but not on other 
proxies of tenant X\footnote{Note that some collateral damage may be
unavoidable if the botnet targets the LB.}.
Active clients of proxies aging-out will redirect to the LB. 
Note that it's possible that redirection could be initiated by the aging-out proxy, depending on the application.
Inactive proxy or LB identities time-out and are then blackholed for a long time (longer for proxy identities).
It's possible that a client's replica does not change during redirection (\eg AWS elastic IP), again depending on the application.
See Figure \ref{fig:motag-p2p}.

Redirecting persistent HTTP connections \cite{RFC2616}
via the HTTP REDIRECT command requires tearing down the current TCP connection and establishing a new one between the client and the new server, which can result in some service interruptions. 
As another example, redirecting streamed-media sessions under RTSP \cite{RFC2326} requires an RTSP TEARDOWN and RTSP SETUP to establish the new RTSP session with the new server.  Additionally,  the control and streaming connections over UDP (RTCP and RTP connections respectively) need to be setup with the new server\footnote{Now, the great majority of online streaming video is over HTTP.}.

The tenant's coordination server can also selectively or randomly
decide which individual proxy to change, \eg
 by selecting those experiencing a sufficiently high number of half-opened TCP sessions and so are likely being actively reconnoitered.
Alternatively, all proxies can change simultaneously and periodically.

In the following, we assume that a subpopulation of
bots engages in 
continual, active reconnaissance of the current 
proxy identities  and then targets the rest of the botnet
prior to launching the DDoS attack.
Again, this assumption is well motivated when the majority of
(IoT based) bots cannot cope with redirection.
Note that botnets are known to
periodically observe the target during an attack
to assess its impact.

{\bf Whether proxy-to-proxy redirection:}
Suppose there is no proxy-to-proxy redirection so that an ``aged-out" (not ``current") proxy no longer takes new sessions and can be deactivated/removed once all of its sessions time out.
So, more active proxies are maintained at greater cost than under proxy-to-proxy redirection.
This may be a minor additional cost when client sessions are short.

{\bf Anonymous reconnaissance:}
Botnet  reconnaissance of LBs or proxies could involve TOR for anonymity and increased apparent source diversity, but TOR is often used for vulnerability scanning and its exit routers may be blacklisted.
A botnet could also use more costly VPNs for anonymity, but they don't give increased source diversity from the attack-target's point of view.
Increased source diversity could also be achieved behind dynamic DNS service.

{\bf Previous related work:}
In \cite{Carroll14,Crouse14},
similar moving-target defenses involved shuffling server addresses.
For a system without proxies, they focus on detecting a
vulnerable server (\ie vulnerability probing not
reconnaissance to identify all servers).  In their threat model,
the attacker essentially samples servers ``with replacement" 
leading to a simple binomial-distribution model;
see Section \ref{sec:cc} and the Appendix for models based on 
Stirling numbers of the second kind.

{\bf Detection of active reconnaissance.}
The proxies may detect active reconnaissance by standard means, \eg by counting over a sliding time-window the number of 
half-opened TCP sessions,
failed login attempts, and/or
successful log-ins/sessions by a single client 
as in a client reputation system.
A client reputation system can also be informed by detected mock sessions (after successful log in).
However, a client reputation system would be undermined
by the attacker's use of anonymity services.

\subsection{Summary increased attacker workfactors}

{\bf Targeting the LBs:}
Under LB motag, a flooding bot needs to periodically query DNS to ensure its attack is not targeting a now blackholded IP.
But frequent queries by the same source IP for the same domain name is an attack signature that is easy to detect.
A smaller number of randomly selected more powerful bots can query 
DNS for the {\em current} LB IPs, possibly via an anonymization service,
and once the current LB IPs have been captured, 
the rest of the botnet is targeted.
If the flooding bots target the LBs and the LBs are overwhelmed, the tenant can autoscale the number of LBs.

{\bf Targeting the proxies:}
If the LB is (or LBs are) not overwhelmed, the bots could simply retarget their flood to the proxies when redirected.
But some IoT-based bots may not be able to cope with the LB-to-proxy or proxy-to-proxy redirection (even if redirection is by IP + port number).
Fewer, powerful bots may be able to mount long realistic sessions and follow redirection, but 
such sessions may be detected or timed-out, and 
a periodic retargeting phase is still needed. 
Alternatively, a subpopulation of bots can repeatedly query DNS for current LBs and retarget the rest of the botnet by collected proxies' (IP,  port number) identities.
This kind of botnet reconnaissance is the focus of the following.
As for LBs, the target can be further diffused
by reactive autoscaling the number of proxies.

\subsection{Summary overhead on the nominal clients}\label{sec:overhead}

Nominal clients need to deal with the
overhead and quality of service degradation
associated with redirection, possibly including additional 
DNS transactions. 
For example,
nominal clients of streaming media service, including live video broadcast, may not experience significant service degradation because their (initially primed) playout buffers can tolerate transmission interruptions due to redirection.
 Clients of more interactive sessions, especially real-time sessions whose dataplane travels through the cloud\footnote{As in online gaming 
managed through a cloud-based server. Note that the dataplane of
video and audio conferencing sessions like Skype 
typically is peer-to-peer between
end-users and the cloud is typically
only used for session initiation and monitoring.},
 may experience some service interruption due to redirection.

To illustrate this, consider the video clips we have posted here
\cite{RTSP-redirect-demo}
for the case of RTSP redirection \cite{RFC2326} of video streaming servers\footnote{We added RTSP redirection and a playout buffer to the
media player available here
\cite{RTSP-Client-Server}.}. 
The clips were created by simply running two servers (port numbers 1051 or 1052
indicated at bottom of the display) and a client video player, all on a single
desktop computer. A different clip is shown for cases with or
without a client playout buffer, the former modeling streaming
and latter modeling interactive video (as in gaming).
The server changes are indicated by  
port changes (``STREAMING FROM"): $1051 \rightarrow 1052$
or $1052 \rightarrow 1051$.
Video playback noticeably stalls longer upon redirection without
a playout buffer than with a 
200-frame (5 second) playout buffer.
For this and other types of interactive applications, 
customized techniques 
have been developed (\eg predictive prefetching)
to try to improve user experience during redirection.

\subsection{Summary direct tenant costs of motag defense}

Regarding the tenant's costs, consider again the ``baseline" tenant
of Figure \ref{fig:tenant-baseline}.
Compared to the baseline tenant, the proposed motag defense has additional
costs for the proxy VMs themselves and for reactive autoscaling.
Autoscaling proxies may be  cheaper than LBs (LBs may be mounted on
 physical routers) and replicas (replicas would need larger VMs for computationally intensive applications).
Obviously, motag defense has costs associated with
swapping one proxy VM (or LB) for another. Note that
external network IO costs associated with replicas of the
baseline tenant are instead borne by proxies under motag defense.

\section{Coupon-collector model of botnet reconnaissance
and moving-target defense}\label{sec:cc}

We relate
the problem of botnet 
 reconnaissance of the current set of moving-target proxies
to an ``adversarial" coupon-collector problem where
\begin{itemize}
\item the adversary is the coupon collector,
\item coupons selection (at mean rate $\beta$)
correspond to probing bots assigned to current proxies, and
\item each proxy server corresponds to a different coupon type ($m$ in total).
\end{itemize}
It's the objective of the collector (botnet) to select coupons so that
one of every current type is obtained before the coupon types are
changed (at mean rate $\delta$).

Assume that the length of
time between successive changes in the $m$ {\em types} of coupons
(proxies)
is exponentially distributed with mean $1/\delta$, 
a memoryless distribution with
high variance  (which 
may be advantageous since a deterministic time may be 
more easily learned, and anticipated, by the attacker).
So, all $m$ coupon  types
are periodically changed according to an independent
Poisson process with rate $\delta$. 

Consider a coupon collector (probing botnet) that selects coupons from the
set of $m$ coupon types uniformly at random\footnote{Note that under 
round-robin assignment of clients to proxies by the LB,
session establishment by nominal clients also needs to be considered.
But if the rate of 
nominal client sessions is relatively significant, proxy assignment
will appear to be randomized from the botnet's point of view.}
(with replacement) at aggregate rate $\beta$.  When the $m$ coupon types
all change, the collector starts over with zero.

Let $\rho=\beta/\delta$.

\begin{proposition}\label{prop:constant}
Under deterministically periodic coupon-selection and
Poisson coupon-type changing processes,
the stationary mean number of
different currently valid coupons obtained is
$$1/(\mbox{e}^{1/\rho}-(1-1/m)).$$ 
\end{proposition}

\begin{proposition}\label{prop:poisson}
Under independent Poisson coupon-selection (rate $\beta$) and
coupon-type changing (rate $\delta$) processes,
the stationary 
mean number of 
different currently valid coupons obtained 
(from a maximum of $m$) is
$$m\frac{\rho}{m+\rho}
=m\frac{\beta}{m\delta+\beta}.$$
\end{proposition}

Note that these two results are close for large $\rho$ since
$\mbox{e}^{1/\rho}\sim 1+1/\rho$ as $\rho\rightarrow \infty$.

The proofs of Propositions  \ref{prop:constant} and \ref{prop:poisson} 
and additional related results on adversarial coupon collection
are given in 
the Appendix.
These results represent two extremes in the way the botnet may probe for 
currently valid proxy identities  - 
inter-probe times that are exponentially distributed (high variance) or
constant-rate probing (zero variance),
\cf comparable figures between Tables \ref{table:compare} and \ref{table:compare2}.

We next show how accurate such models are with regard to
our AWS emulations and use them to extrapolate to larger-scale scenarios.

\section{Performance of AWS prototype}\label{sec:emul}

\subsection{Experimental set-up}

On AWS, we set-up
HTTP sessions between users and replicas -
specifically, Apache web servers with a PHP shopping cart application that 
interacts with a MySQL database (essentially the LAMP stack).
All proxies reside in containers that
are spun up ahead of planned proxy address change (again, this may
not be needed 
if a service like AWS elastic IP is used). 

For our emulation results described below, 
the mean time between probes by each reconnoitering bot was 30s, and
the  
time between proxy-identity  changes (all proxies change identities
and at the same times)
had mean 30s.
Assignment of proxies to bots is either randomized or round-robin,
the latter requiring emulation of the session setup process of nominal clients.

\subsection{Emulation results}\label{sec:AWS-results}

\begin{figure}[!htb]
\begin{center}
\ifTR
\includegraphics[width=0.75\textwidth]{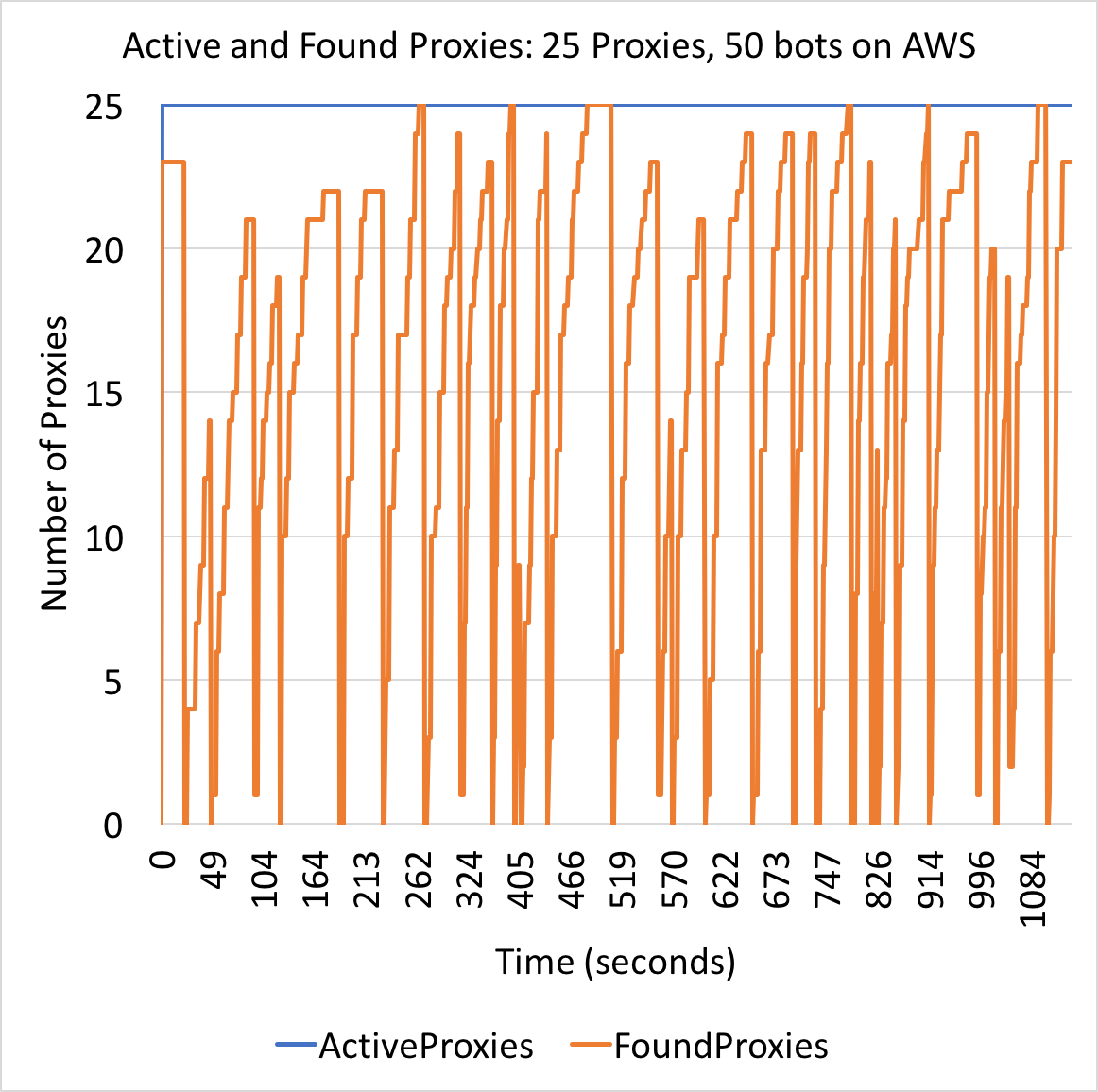}
\else
\includegraphics[width=0.8\columnwidth]{./figure-aws/M25_z50.png}
\fi
\caption{Number of current proxies identities known to the
botnet over time with  $m=25$, $\rho=50$.}\label{fig:AWS-M25_z50}
\end{center}
\end{figure}

\begin{figure}[!htb]
\begin{center}
\ifTR
\includegraphics[width=0.75\textwidth]{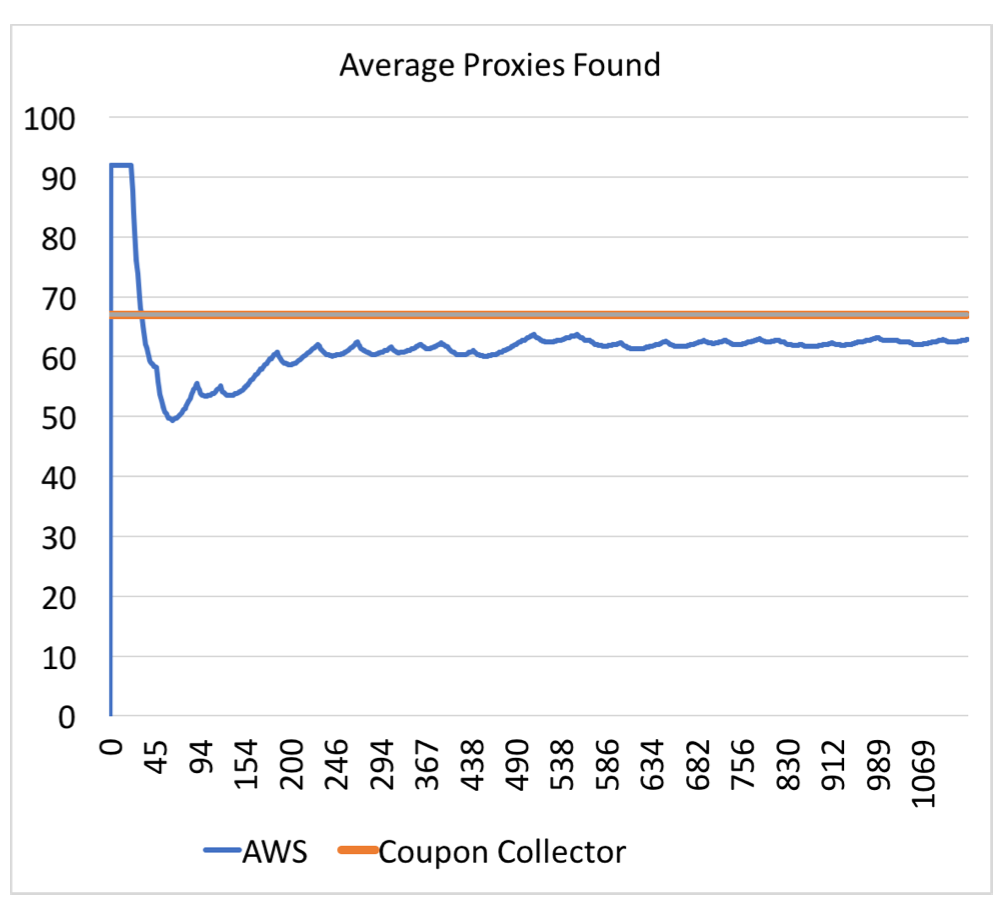}
\else
\includegraphics[width=0.8\columnwidth]{./figure-aws/M25_z50_avgb.png}
\fi
\caption{Running time-average number of current proxies identities known to the
botnet with $m=25$,  $\rho=50$.}\label{fig:AWS-M25_z50_avg}
\end{center}
\end{figure}

A typical experimental result from our AWS emulations is
given in Figs. \ref{fig:AWS-M25_z50} and \ref{fig:AWS-M25_z50_avg}
for: $m=25$ proxies; $50$ bots each with mean of the
exponentially distributed inter-probe time
$30$ seconds so that the 
total mean probing rate was $\beta=50/30$;
and the mean of the exponentially distributed
time between successive proxy identity changes
is $30$ seconds so $\delta =1/30$.
Thus, $\rho=\beta/\delta = 50$. 
Note that, generally, $\beta$ (and hence $\rho$)
 may be roughly proportional to the number of
probing bots.
Finally, randomly chosen proxies are assigned to probes from the bots.

Table \ref{table:compare} compares the results of 
the AWS emulations, coupon-collector model of Prop. \ref{prop:poisson}. 
\begin{table}
\begin{center}
\begin{tabular}{|c||c|c|c|}
\hline
method ~$\backslash$~ $m,\rho$
&  25,5 & 25,25 & 25,50 \\ \hline\hline
AWS emulations& 16\%  & 48\% & 62\% \\ \hline
coupon collector (Prop. \ref{prop:poisson}) & 17\% & 50\% & 67\%   \\ \hline
\end{tabular}
\caption{Stationary mean percentage of proxies known to the botnet
under Poisson proxy changes and probing}\label{table:compare}
\end{center}
\end{table}

\begin{figure}[!htb]
\begin{center}
\ifTR
\includegraphics[width=0.75\textwidth]{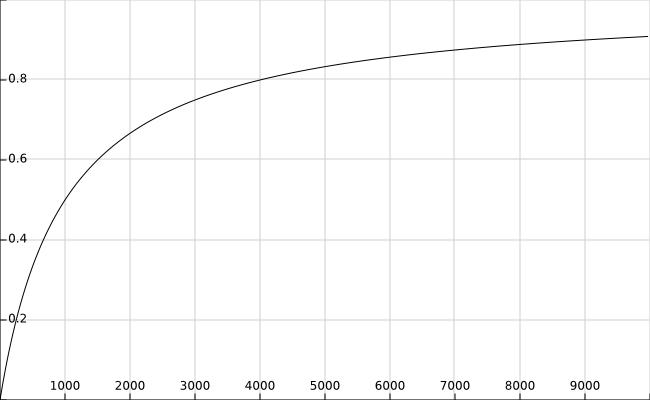}
\else
\includegraphics[width=\columnwidth]{./figure-aws/M1000-cc.png}
\fi
\caption{The average fraction of proxies known
to the botnet as a function of $\rho=\beta/\delta$ 
for $m=1000$ proxies.}\label{fig:M1000-cc}
\end{center}
\end{figure}

\begin{figure}[!htb]
\begin{center}
\ifTR
\includegraphics[width = 0.8\textwidth]{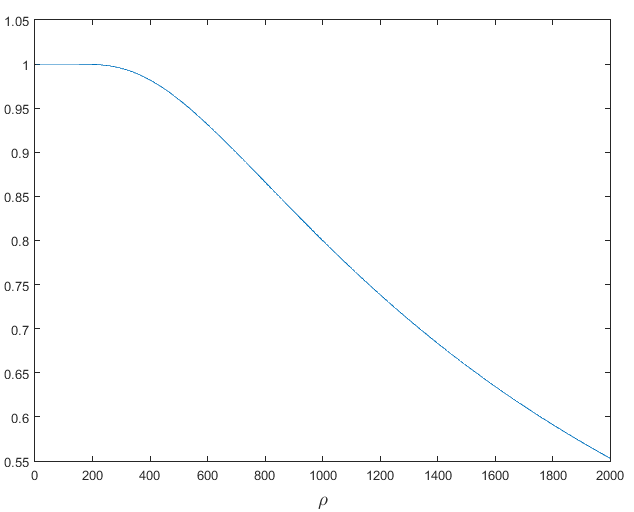}
\else
\includegraphics[width = \columnwidth]{./figure-arxiv/geq20uncompromised1000.png}
\fi
\caption{The probability that at least 20\% of proxies are not
found for $m=1000$ servers under Poisson
probing.}\label{fig:geq20uncompromised}
\end{center}
\end{figure}

In Fig. \ref{fig:M1000-cc}, we use the coupon collector
model of Prop. \ref{prop:poisson} to extrapolate to the case of $m=1000$ servers.
In Figure \ref{fig:geq20uncompromised}, we use the distribution
for the Poisson (exponential) probing case 
(see Proposition \ref{prop:distrn-poisson} in the Appendix)
to plot the probability that at least 20\% of the servers
are not found as a function of $\rho$.

\begin{table}
\begin{center}
\begin{tabular}{|c||c|c|c|}
\hline
method ~$\backslash$ ~$\kappa$ 
&  .05   &.25 & .5 
\\ \hline\hline
AWS emulations& 67.3\% & 66.8\%  & 54.6 \% 
\\ \hline
\end{tabular}
\caption{Mean percentage of proxies known to
the botnet under deterministic probing with $m=25,\rho=50$
under ``truncated" Gaussian interprobing times
}\label{table:compare2}
\end{center}
\end{table}

For Table \ref{table:compare2}, 
we instead use truncated Gaussian distributed inter-probe  times
lower bounded by 2 seconds.
Here, the variance of inter-probe times is taken
smaller than under the exponential distribution considered above.
The low-variance case $\kappa=0.05-0.25$ is very close to 
67.1\% given by Prop. \ref{prop:constant} for deterministic-rate probing.
To see why,
note that because $\beta \gg \delta$  ($\rho\gg 1$) in this case,
there are typically many probes over the time $T$
between
proxy address changes 
 ($\E T=1/\delta$);
so, with low-variance inter-probe times, 
by the law of large numbers one would continue to expect
the number of probes to likely be about $\beta T$ 
as for constant probing,  
see the Appendix.
About $2/3$ of the proxies are
known to the botnet on average 
in steady-state 
(\ie about $1/3$ would be available for nominal clients during the attack)
when $m=25,\rho=50$ for both deterministic and Poisson probing,
which is a lot greater than under higher-variance Gaussian probing,
\cf 
(\ref{clt}) of the Appendix.
Other relevant models of coupon collection are considered in 
the Appendix,
\eg where individual proxies change independently rather
than collectively.

\section{Summary}\label{sec:concl}

In this paper, we considered a botnet intending to launch
at DDoS attack against a tenant of the public cloud.
We motivated how a tenant whose clients are
redirected to a bank of proxy servers 
that periodically change compels a botnet to actively probe 
the tenant to determine a set of currently valid proxy identities
for targeting purposes.
The performance of this moving-target defense against botnet reconnaissance was studied by AWS emulations and by adversarial coupon-collection 
models to show
how, at any given time, a certain number of proxies are not known to the botnet
and would be available to nominal clients during the DDoS attack.
Those servers will continue to be available to the community of nominal clients even after the rest of the botnet is targeted and
the DDoS attack is launched.

{\footnotesize 
\bibliographystyle{acm}
\bibliography{./bibfiles/neda,./bibfiles/stirling,./bibfiles/ddos,./bibfiles/cloud,./bibfiles/p2p,./bibfiles/unstruct,./bibfiles/stochastic,./bibfiles/IoT,./bibfiles/ref_cheng}
}

\setcounter{proposition}{0}
\input{stirling3-appendix.tex}

\end{document}

%% file: define.tex
\def\E{{\sf E}}
\def\P{{\sf P}}

\def\eg{{\em e.g.},~}
\def\ie{{\em i.e.},~}
\def\cf{{\em cf.},~}
\newcommand{\beqa}{\begin{eqnarray*}}
\newcommand{\eeqa}{\end{eqnarray*}}
\newcommand{\be}{\begin{eqnarray}}
\newcommand{\ee}{\end{eqnarray}}
\newtheorem{proposition}{Proposition}

%% file: stirling3-appendix.tex
\section{Appendix: Adversarial coupon-collector models}
We consider  a coupon-collector problem involving an adversary.
In a cyber security interpretation,
the collector  is the attacker targeting
a bank of servers and its adversary is the defender.
That is, suppose bots of a bot network (botnet) 
are conducting reconnaissance on 
the servers by selecting one at random and communicating with
it in order to identify it. The servers protect against reconnaissance
by  periodically changing their identities.  
So that the botnet cannot anticipate when server identities change,
we assume that times between these events are 
independent  and (memorylessly) exponentially distributed.
Thus the bots need to continually sample the servers to
capture their current identities.

In terms of coupon collection, assume that there is
a coupon collection process and that,
periodically, the coupon types are replaced so that the
collector needs to start over.  There are $m$ different
types of coupons independently chosen uniformly at
random. Let $Y\leq m$ be the process representing
the number of different coupon types possessed by the collector and
$K$ (or $K(T)$) be the number of coupons selected in the interval $T$ between
coupon-type replacements.
We develop continuous-time models in the following 
using Stirling numbers of the second kind
and an associated distribution.
In Table \ref{table:jargon}, we give the correspondence between
this cyber security and the coupon collector model.

\begin{table}[h!]
\centering
\begin{tabular}{|c|c|c|}
\hline
cyber defense context & coupon collector context & parameter \\\hline\hline
reconnaissance attack & coupon selection & rate $\beta$ \\\hline
server/target & coupon type & $m$ types \\\hline
defense & change coupon type & rate $\delta$ \\\hline
\end{tabular}
\caption{Cyber defense and coupon collector jargon correspondence}\label{table:jargon}
\end{table}

This Appendix is organized as follows.
In  Section
\ref{sec:back}, we give some background on Stirling
numbers of the second kind. 
In Section \ref{sec:replace-all}, we consider models where
periodically all coupon types are changed at the same time.
In Section \ref{sec:replace-selectively}, we consider models where
periodically coupon types are selectively replaced independently.

\subsection{Background on Stirling Numbers of the Second Kind}\label{sec:back}

Given coupons of $m$ different types.
If $k$ distinct coupons are independently selected,
the number of ways of selecting 
$y\leq k\wedge m :=\min\{k,m\}$ 
distinct types is
\beqa
\frac{m!}{(m-y)!}{k \brace r},  & \mbox{where} &
{k \brace y}  =  \frac{1}{y!}\sum_{j=0}^y (-1)^{y-j}\binom{y}{j}j^k
\eeqa
is a Stirling number of the second kind.
Assuming that the type of a selected coupon is equally likely
any of the $m$ types,
the  probability that $k$ distinct coupons
are selected so that $y$ of $m$ distinct types result is
\be\label{sdsk}
P_{m,k}(y)  & := &  \frac{\frac{m!}{(m-y)!}{k \brace y}}{m^k}  
\ee
where
\beqa
m^k    &= &  \sum_{y=1}^{k\wedge m}
\frac{m!}{(m-y)!}{k \brace y} ,~
y\in\{1,2,...,k\wedge m\},
\eeqa
see Equation (1.55) of \cite{Johnson-Kemp-Kotz}.
Note that this discrete distribution $P_{m,k}$
is different from the ``Stirling distribution of the second
kind" (Equation (4.98) of \cite{Johnson-Kemp-Kotz}).

From, \eg  (1.94c) on p. 82 of \cite{Stanley11}, we have that
the power series 
\be
\sum_{k=0}^\infty {k \brace \ell} x^{k+1} = \frac{1}{(1/x)_{\ell+1}},
\label{power-series}
\ee
where $(a)_\ell := a(a-1)(a-2)\cdots(a-\ell+1)$.

For numerical computations, 
we can employ known asymptotic approximations
of the
Stirling numbers of the second 
kind. For example,
if $\ell$ is fixed then
${k \brace \ell} \sim \ell^k/\ell!$ 
as $k\rightarrow \infty$, 
see 26.8.42 at \cite{dlmf.nist} (also see
\cite{Temme93,LOU}).
Then using logarithms, we 
can compute  the distributions
$P_{m,k}$. 
When $k\gg m$ or $k\ll m$,  the distribution $P_{m,k}(\cdot)$
is concentrated at its mode very near $k\wedge m$.
However when $k=m$,  the mode 
is about 20-25\% less than the maximum
(at $k=m$).

\subsection{Periodically replacing all coupon types}\label{sec:replace-all}

\subsubsection{Deterministic coupon selections with Poisson coupon-type replacements}

\begin{proposition}
Under constant rate coupon-selection (constant rate $\beta$),
and Poisson coupon-changing process (mean rate $\delta$) 
determining when all coupon types are replaced,
the stationary mean number of 
currently valid, different coupon types obtained  is
\be
\E Y &  = & \frac{1}{\mbox{e}^{1/\rho}-(1-\frac{1}{m})}. 
\label{cc-constant}
\ee
\end{proposition}

{\bf Proof:}
At constant $\beta$ selections/s,  the number of 
coupons selected over $[0,T]$ is simply 
\beqa
K & = &[\beta T]. 
\eeqa
$K$ is a geometric random variable,
\[
\P(K=k) = \P(k/\beta \le T < (k+1)/\beta)
= \mbox{e}^{-k/\rho}\bigg(1-\mbox{e}^{-1/\rho}\bigg), 
\]
for $k=0,1,2\ldots$,
with probability generating function
\[
\E z^{K} = \frac{1-\mbox{e}^{-1/\rho}}{1-z \mbox{e}^{-1/\rho}}.
\]
So,
\[
\E Y =
\E m(1- (1-\frac{1}{m})^K) 
= \frac{\mbox{e}^{-1/\rho}}{1-(1-\frac{1}{m}) \mbox{e}^{-1/\rho}} ~,
\]
which is (\ref{cc-constant}).
Since Poisson Arrivals See Time Averages
(PASTA) \cite{Wolff89,Bremaud91}, this is also the expected number of different
types of coupons collected at a typical time.
\qed

\subsubsection{Poisson coupon selection with independent Poisson coupon-type replacement process}

\begin{proposition}
Under Poisson coupon-selection (rate $\beta$),
Poisson coupon-changing process (rate $\delta$),
the stationary mean number of 
currently valid, different coupon types obtained  is
\be
\E Y & = & m\frac{\rho}{m+\rho}. \label{cc}
\ee
\end{proposition}

{\bf Proof:}
The number of coupon types collected just before coupon types
are replaced is
\beqa
\E m(1-(1-1/m)^K).
\eeqa
Conditioning on $T$,
$K\sim{\sf Poisson}(\beta T)$.  
So, 
the mean number of different coupons obtained conditioned on $T$ 
is simply
\be
\sum_{k=0}^\infty  m(1-(1-1/m)^k)\frac{(\beta T)^k \mbox{e}^{-\beta T}}{k!} 
 = 
m(1-\mbox{e}^{-\beta T/m}). \label{Poisson-probing}
\ee
Using the moment generating function of $T\sim\exp(\delta)$, we get
\beqa
\E m(1-(1-1/m)^K) & =&\E m(1-\mbox{e}^{-\beta T/m})\\
& = & m\left(1-\frac{\delta}{\delta+\beta/m}\right)
\eeqa
which is (\ref{cc}).
Finally, since PASTA,
this is also the expected number of different
coupons collected at a typical time.
\qed

~\\

Note that (\ref{cc}) and (\ref{cc-constant}) are close
when $\rho$ is large since $\mbox{e}^{1/\rho} \sim 1+1/\rho$.

Also note that when $m\gg \rho$, the expected number 
of different coupons in steady state is approximately $\rho$ 
which  has nothing to do with $m$. On the other hand,
if $\rho \gg m$, all $m$ coupons are likely collected following intuition.

\begin{proposition}\label{prop:distrn-poisson}
Under Poisson coupon-selection (rate $\beta$) and
Poisson coupon-changing process (rate $\delta$),
in steady state, 
\[
\P(Y=\ell)  =  \frac{1}{\rho}\, \frac{m}{m-\ell}\,\,
\frac{m}{m \frac{\rho+1}{\rho}}\, \frac{m-1}{m \frac{\rho+1}{\rho}-1} \cdots 
\frac{m-\ell}{m \frac{\rho+1}{\rho}-\ell}, 
\]
for $1\le \ell \le m$.
\end{proposition}

{\bf Proof:}
Again, $K$ given $T$ is $\sim{\sf Poisson}(\beta T)$.
The unconditional distribution of $K$ is geometric:
\be
\P(K(T)=k) & = &  \int_0^\infty \delta \mbox{e}^{-\delta t} 
\frac{(\beta t)^k\mbox{e}^{-\beta t}}{k!} \mbox{d}t \nonumber \\ 
& = & \bigg(\frac{\rho}{\rho+1}\bigg)^k \, \frac{1}{\rho+1},
\label{K_T-geom}
\quad k \ge 0 
\ee
Thus, for $\ell \in \{0, 1,...,m\}$,
\beqa
\P(Y(K(T))=\ell) 
& =&  \sum_{k=0}^\infty \P(Y(k)=\ell) \bigg(\frac{\rho}{\rho+1}\bigg)^k \frac{1}{\rho+1}
\\
& =&  \sum_{k=0}^\infty  \frac{(m)_\ell}{m^k} {k \brace \ell} \bigg(\frac{\rho}{\rho+1}\bigg)^k \frac{1}{\rho+1}
\\
& =&  \frac{(m)_\ell}{\rho+1} \frac{\rho+1}{\rho/m}
\sum_{k=0}^\infty  {k \brace \ell} \bigg(\frac{\rho/m}{\rho+1}\bigg)^{k+1}
\\
& =& \frac{m}{\rho}\, (m)_\ell \frac{1}{\bigg(\frac{\rho+1}{\rho/m}\bigg)_{\ell+1}}
\eeqa
where the last equality is (\ref{power-series}).
Again since PASTA, this is the steady-state distribution
of the number of coupon types obtained.
\qed

~\\

See the Markov chain below with transition rates (\ref{rate-up}) and
(\ref{rate-down})  and parameter $r=1$.

\subsubsection{Discussion: Renewal coupon selections, Poisson coupon-type replacements}\label{sec:CLT}

Suppose that the coupon selection process $K(t)$ is a renewal 
counting process
with interarrival times $X$ of finite variance, $\sigma^2<\infty$,
and mean $1/\beta$.
The classical central limit theorem for renewal processes is
(\eg \cite{renewal-gallager}),
\beqa
\lim_{t\rightarrow\infty}\P\left(\frac{K(t)-\beta t}{\sqrt{\sigma^2\beta^3 t}} 
< y\right) & = & 
\frac{1}{\sqrt{2\pi}}\int_{-\infty}^y \mbox{e}^{-z^2/2}\mbox{d}z,
\eeqa
\ie 
$K(t)\sim {\sf N}(\beta t, \sigma^2 \beta^3 t)$ as $t\rightarrow \infty$.
We can consider a regime where 
$\delta^{-1} \gg \beta^{-1}$  ($\rho \gg 1$)
so that this is approximately
valid for $K(T)$ when $T\sim \exp(\delta)$. 
Let $v=-\log(1-1/m)>0$.
Using the MGF of a Gaussian, we can approximate the expected
number of different coupon types collected just before they're replaced 
as:
\be
 \E m(1-(1-1/m)^{K(T)}) & \sim & 
\E m(1-\mbox{e}^{-\beta T v + \sigma^2 \beta^3 T v^2/2}) \nonumber \\
 & & ~~~~\mbox{as $\rho=\beta/\delta \rightarrow \infty$} \label{clt}\\
& = & m \left(1- \frac{\delta}{\delta + \beta v - \sigma^2 \beta^3 v^2/2}\right)
\nonumber
\ee
requiring $\delta + \beta v - \sigma^2 \beta^3 v^2/2 > \delta $,
\ie $\beta^{-1}>\sigma\sqrt{v/2}$.
That is, when $\rho=\beta/\delta$ is large, there are obviously many
coupons selected before the coupon types are changed.
Again since PASTA, this is the stationary number of 
different coupon-types collected.
When $\sigma=0$,
this limit agrees with the deterministic coupon selection
(\ref{cc-constant})
of Proposition \ref{prop:constant}
when $m$ and $\rho$ are large since $v\sim 1/m$
and $\mbox{e}^{1/\rho}\sim 1+1/\rho$.

\subsection{Individual coupon types 
independently replaced at random, according to a
single Poisson process}\label{sec:replace-selectively}

Consider the discrete-time Markov chain $Y(k)$ satisfying
\beqa
Y(k) & = & Y(k-1) + \xi_k, \quad k \ge 1,
\eeqa
where  $Y(0)\in \{0,1,2,...,m\}$ (\ie $Y(0)$ is not necessarily zero a.s.) and
\beqa
\P(\xi_k = 0 | Y(k-1),\ldots, Y(0)) & = & \frac{Y(k-1)}{m}, \\
\P(\xi_k = 1 | Y(k-1),\ldots, Y(0)) & = &1-\frac{Y(k-1)}{m}.
\eeqa
Here, $Y(k)$ is the number of different types of coupons collected
after the $k^{\rm th}$  coupon selection including those initially
had, $Y(0)$.

\subsubsection{Verifying (\ref{cc}) when $Y(0)=0$ a.s.}
We can alternatively verify (\ref{cc}) by simply solving 
the recursion for $\E Y(k)$ when $Y(0)=0$:
\beqa
\E Y(k) & = & 1+(1-\frac1m)+(1-\frac1m)^2+\cdots+(1-\frac1m)^{k-1}\\
 & = & m\bigg(1-(1-\frac1m)^k\bigg).
\eeqa
Note that $Y(0)=0$ a.s. implies $\xi_1=1$ a.s.
Now let us collect coupons at the ticks of a Poisson process $K(t)$, $t\ge 0$,
 with rate $\beta$ and independent coupon types selected independently and
uniformly.
Thus, at real time $t \ge 0$, we have collected
\[
Y(K(t))~ \text{ different types of coupons}.
\]
So,
\beqa
\E(Y(K(t))) & = & \sum_{k=0}^\infty m\bigg(1-(1-\frac1m)^k\bigg) \P(K(t)=k)\\
& = & m(1-e^{-\beta t/m}).
\eeqa
So if $t=T$ is exponentially distributed  with rate $\delta$,
\beqa
\E(Y(K(T))) & =& \int_0^\infty m(1-e^{-\beta t/m}) \delta e^{-\delta t} \mbox{d}t\\
& = & \frac{m\rho}{m+\rho}. 
\eeqa

\subsubsection{Independent coupon-type replacements, $Y(0)\not\equiv 0$ a.s.}

A generalization of (\ref{cc}) is:
 
\begin{proposition}\label{prop:mean-r}
For Poisson coupon selection (at rate $\beta$),
Poisson coupon-type replacements (at rate $\delta$), and
individual coupon types are independently replaced with
probability $r$, 
the mean number of valid coupon types
obtained in steady state is
\be
m\frac{\rho}{rm + \rho}
\label{cc-r}
\ee
\end{proposition}

{\bf Proof:}
For $Y(0)\not\equiv 0$ a.s., solving the recursion 
\beqa
\E Y(k) & = & \E Y(k-1)-\E \xi_k ~=~ 1+(1-1/m)\E Y(k-1)
\eeqa
gives, for $k\geq 1$,
\beqa
\E Y(k)  & = & 1+(1-1/m)+...\\
 & & ~~~ ... +(1-1/m)^{k-1}+(1-1/m)^k \E Y(0)\\
 & = & m(1-(1-1/m)^k) + (1-1/m)^k \E Y(0).
\eeqa
As above, since $K\sim {\sf Poisson}(\beta T)$ and
$T\sim \exp(\delta)$,
\beqa
\E Y(K(T))  
 & = & m\frac{\rho}{m+\rho} + \frac{m}{m+\rho} \E Y(0).
\eeqa
In steady-state,
\be
(1-r)\E Y(K(T)) & = & \E Y(0). \label{mean-r}
\ee
Eliminating $\E Y(0)$ from the previous two displays gives
$\E Y(K(T))$ equals (\ref{cc-r}).
The proposition follows since PASTA.
\qed

~\\

Note that when $r=0$ (coupon types are never replaced), 
the mean number of coupon types obtained in steady-state is $m$.

Here we are modeling the number of coupon types collected as
a Markov chain with the following transition rates:
\be
q(k,k+1) & = & \frac{m-k}{m}\beta, ~~0\leq k\leq m \label{rate-up}\\
q(\ell,k) & = & \delta \binom{\ell}{k} r^{\ell-k}(1-r)^k,~
0\leq k<\ell \leq m  \label{rate-down}
\ee
One can find the stationary invariant distribution $\pi$ through the balance
equations, $\pi^{\rm T}q =0$.
For $r=1$, this stationary invariant  $\pi$ 
is that of Proposition \ref{prop:distrn-poisson}.

\subsubsection{Individual coupon types replaced according to
independent Poisson processes}

A simpler variation is one in which 
each coupon  is reset according to 
an {\em independent} Poisson process at rate $\delta$.
Here, the upward transition rates are (\ref{rate-up}) but, 
instead of (\ref{rate-down}), the downward transition rates  are
\beqa
q(\ell,\ell-1)= \delta \ell, ~~ 1\leq \ell \leq m.
\eeqa
So, this is a birth-death Markov process.
Here, a separate $\sim \exp(\delta)$ clock would be 
required to reset each coupon type.
The solution of the balance equations is the invariant
$\pi$ on $\{0,1,2,...,m\}$:
\beqa
\pi(k)& = & \pi(0)\prod_{j=1}^k \frac{q(j-1,j)}{q(j,j-1)} \\
& = & \pi(0) \binom{m}{k}
\left(\frac{\rho}{m}\right)^k ,
\eeqa
where $\pi(0)$ is chosen so that $\sum_{k=0}^m \pi(k)=1.$